\documentclass[final]{aipproc}

\layoutstyle{6x9}

\begin{document}

\title{The Electron-Ion Collider}


\author{V. Guzey}{
  address={Theory Center, Thomas Jefferson National Accelerator Facility, 
Newport News, VA 23606, USA}, email={vguzey@jlab.org}}

\begin{abstract}

The future Electron-Ion Collider (EIC) is a proposed new facility to
collide high-energy electrons with beams of polarized protons/light nuclei and 
unpolarized nuclei. We overview the goals of the project and key measurements at the EIC.
We also briefly comment on recent developments of the project.

\end{abstract}

\maketitle


\section{Introduction}

One of the key goals of nuclear physics is to understand the structure of
hadrons is terms of quarks and gluons. After decades of investigations, the following
questions remain open:
\begin{itemize}
\item[(i)] What is the gluon momentum distribution in nuclei?
\item[(ii)] What are the properties of high-density gluon matter?
\item[(iii)] How do fast partons interact as they traverse nuclear matter?
\item[(iv)] How do partons contribute to the spin structure of the nucleon?
\item[(v)] What is the spacial distribution of partons in the nucleon?
\item[(vi)] How do hadronic final states form in QCD?
\end{itemize}
The Electron-Ion Collider (EIC) is the proposed new facility to collide high-energy
electrons with nuclei and polarized protons/light nuclei, which will address the open questions just mentioned~\cite{Deshpande:2005wd,eic,eic_general,eic_eA,meic}. 
Two broad classes of goals of the future EIC are reflected in two
physics working groups (WG) of the EIC collaboration: the eA WG concentrates 
on exploring the (strong) gluon fields in nuclei, and the ep WG focuses on the precision
imaging of quarks and gluons in the nucleon.

The planned physics program of the future EIC imposes certain requirements on the EIC.
These include: a lepton beam that provides a clean and well-understood probe;
high and variable c.m.~energy ($\sqrt{s}=50-90$ GeV and possibly lower)
that will allow to have large $x-Q^2$ kinematic coverage
and access the longitudinal structure function $F_L(x,Q^2)$;
the luminosity ${\cal L} > 10^{33}$~cm$^{-2}$~s$^{-1}$;
a polarized lepton beam to study the nucleon spin structure and a polarized beam of light
nuclei (D and $^3$He) to perform flavor separation of polarized parton distributions (PDFs)
and to study the Bjorken sum rule; unpolarized nuclear beams
(light nuclei will help to perform flavor separation, and heavy nuclei will be used to
study the high-density gluon matter).

\section{Key measurements at EIC}

Below we discuss selected key basic measurements at the EIC.

\subsection{Unpolarized DIS}

The fundamental measurement will be inclusive deep inelastic
scattering (DIS) with protons and nuclei and the measurement of the structure functions $F_2(x,Q^2)$ and $F_L(x,Q^2)$. The structure function $F_2(x,Q^2)$ probes quarks directly and gluons
through the scaling violations; the longitudinal structure function $F_L(x,Q^2)$
will access gluons directly. In addition, in the case of nuclear beams, these 
measurements will probe such nuclear effects as nuclear shadowing and antishadowing
and the EMC effect.

Complimentarily to $F_2(x,Q^2)$ and $F_L(x,Q^2)$, one will measure the charm structure functions $F_2^c(x,Q^2)$ and $F_L^c(x,Q^2)$ through open charm production. They will give
an access to the charm quarks and, again, to gluons.

Another measurement will be the measurement of light and heavy jets 
in DIS. This will give additional constraints on quark and gluon PDFs.
In addition, in the case of nuclear beams, the measurement of jets provides the cleanest environment to study nuclear modifications of hadron production. 
Simultaneous studies of
light and heavy quark jets will test models of hadronization (energy loss vs.~absorption)
in cold nuclear metter, which is relevant for the interpretation of RHIC and LHC data.

\subsection{Diffractive DIS}

At the lepton-proton collider HERA, diffraction in DIS turned out to be one of key
measurements with many new and unexpected results. 

The basic measurement at the EIC will be inclusive diffraction in DIS with protons and
nuclei. First, diffraction in 
DIS with nuclei has never been measured, which will bring us into a completely uncharted territory. Eventually, one hopes to learn about the nature of Pomeron in QCD.
Second, the diffractive structure function $F_2^{D}$ is expected to be more sensitive
to the phenomenon of saturation of parton (gluon) densities than the inclusive structure
function $F_2(x,Q^2)$.

A complimentary measurement will be jets in diffractive DIS, which will place additional
constraints on quark and gluon diffractive PDFs.

Another very important set of measurements will be exclusive diffraction: deeply virtual
Compton scattering (DVCS) and electroproduction of mesons.
DVCS and electroproduction of vector mesons will probe the singlet quark and gluon
generalized parton distributions (GPDs), with the aim to learn about the transverse
imaging of hadrons. In addition, one expects that such nuclear effects as color
transparency, nuclear shadowing and saturation will be enhanced in these measurements.

One should also mention non-diffractive electroproduction of $\pi$, $K$, $\rho^+$, 
etc.~mesons,
which will probe the non-singlet, valence structure of the hadron beam.

\subsection{Polarized DIS}

We will only briefly mention several key measurements that
can be performed at the EIC
with polarized lepton-proton
collisions; please see Ref.~\cite{Deshpande:2005wd} for details.

The basic measurement will be inclusive DIS, which will aim to constrain the gluon polarized
PDF $\Delta g(x,Q^2)$ through the scaling violations, which will be possible because of
the wide kinematic coverage in the $x-Q^2$ plane.
Additionally, one will constrain $\Delta g(x,Q^2)$ from open charm and dijet
production via the process of photon-gluon fusion.

Another important measurement will be semi-inclusive DIS, which will enable one to
perform flavor separation of polarized PDFs and to access transverse momentum dependent
PDFs. 

Also, many exclusive measurements will also be performed in the polarized mode. 

\section{One example: nuclear gluon PDF at EIC}

Nuclear PDFs, especially the gluon nuclear distribution, are poorly known at 
small Bjorken $x$. Nuclear PDFs are extracted from the data on DIS with fixed 
nuclear targets, which have a limited kinematic coverage where small $x$ are either
inaccessible or correspond to low $Q^2$. As a result, 
the small-$x$ extrapolation of the results of global fits to nuclear DIS  has large uncertainties 
and, in general, leads to essentially unconstrained nuclear PDFs. In addition, 
the nuclear gluon PDF is not measured directly, but is extracted through the scaling 
violations.

While nuclear PDFs in the shadowing region are interesting in their own right,
the knowledge of nuclear PDFs at small $x$ is essential for the interpretation of RHIC and future
LHC data (e.g.,~saturation vs. leading-twist mechanisms). Also, precise nuclear PDFs are needed to separate the initial state effects from final state effects (parton energy loss) and to test different
models of fragmentation.

\begin{figure}
\includegraphics[height=.27\textheight]{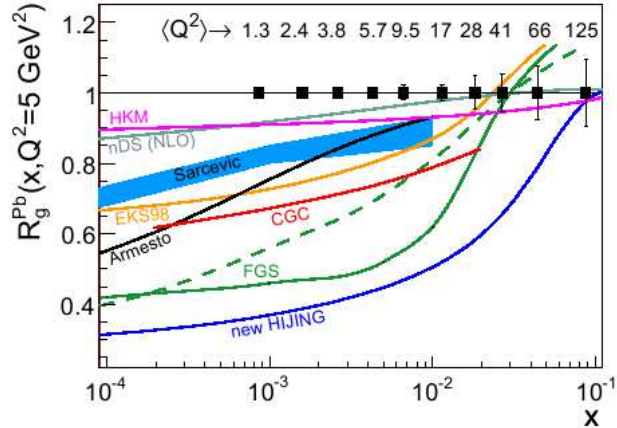}
\caption{The ratio of the gluon distribution in $^{208}$Pb to that in the free proton, $R_g$, see Eq.~(\ref{eq:Rg}), as a function of Bjorken $x$.
Different curves correspond to the extrapolation
of the results of the global fits to the unmeasured small-$x$ region and to several theoretical
models of nuclear shadowing.
The filled squares and corresponding error bars represent the projected kinematic reach 
and statistical uncertainty of the extraction of $R_g$ using the measurement of
$F_L(x,Q^2)$ at the EIC.
}
\label{fig:guzey_Rg}
\end{figure}

The uncertainty in the present knowledge of the nuclear gluon distribution is  demonstrated in Fig.~\ref{fig:guzey_Rg}, where the ratio of the 
gluon distribution in $^{208}$Pb to that in the free proton, $R_g$,
\begin{equation}
R_g(x,Q^2)=\frac{g_A(x,Q^2)}{Ag_N(x,Q^2)} \,,
\label{eq:Rg}
\end{equation}
is plotted as a function of Bjorken $x$. Different curves correspond to the extrapolation
of the results of the global fits to the unmeasured small-$x$ region and to several theoretical
models of nuclear shadowing~\cite{eic}.

The future EIC with a wide kinematic coverage will allow to determine the gluon nuclear
PDF at small $x$ using several complimentary measurements: direct measurement through
the longitudinal structure function $F_L(x,Q^2)$, indirect measurements through the scaling violations of 
$F_2(x,Q^2)$, and also using the charm structure functions $F_2^c(x,Q^2)$ 
and $F_L^c(x,Q^2)$ as well as jet measurements. The filled squares and corresponding error
 bars in Fig.~\ref{fig:guzey_Rg} represent the projected kinematic reach 
and statistical uncertainty of the extraction of $R_g$ using the measurement of
$F_L(x,Q^2)$ at the EIC.

\section{Concepts of the collider and status of project}

The original design of the EIC (ca.~2006) involves two concepts: eRHIC on the base of RHIC, where
an additional energy recovering linac has to added, and ELIC at Jefferson Lab, which
requires a construction of a new hadron facility to be used with the existing CEBAF.
The eRHIC concept allows for larger $\sqrt{s}=60-90$ GeV and smaller luminosity
${\cal L} \approx 10^{33}$ cm$^{-2}$~s$^{-1}$, while the ELIC concept corresponds to smaller
$\sqrt{s} \leq 60$ GeV and larger luminosity ${\cal L} \approx 10^{35}$ cm$^{-2}$~s$^{-1}$.

Both concepts have their merits and shortcomings. The recent development of the EIC project
goes along the lines of the discussion of the so-called staged approach to eRHIC, 
where one starts with
a lower energy eRHIC ($2-4 \times 100$ GeV for the eA mode and $2-4 \times 250$ for the ep mode),
and the low/medium energy EIC at Jefferson Lab ($5 \times 30-60$ GeV)~\cite{meic}. 

Regardless of the eventually chosen concept of the future EIC, the present goal of the EIC collaboration is to strengthen the physics case and to determine the detector and collider designs, which should be 
convincing enough to obtain a high-level recommendation/approval of the entire project by the 
2012 NSAC Long Range Plan.

\bibliographystyle{aipproc}   

\end{document}